\documentclass[conference]{IEEEtran}

\usepackage{cite}
\usepackage{graphicx}
\usepackage[cmex10]{amsmath}
\usepackage{amsfonts,amssymb,amsthm}
\usepackage[english]{babel}
\usepackage{balance} 
\usepackage{xcolor}
\usepackage[normalem]{ulem} 
\usepackage{enumitem}
\usepackage[caption=false, font=footnotesize]{subfig}

\def\BibTeX{{\rm B\kern-.05em{\sc i\kern-.025em b}\kern-.08em
    T\kern-.1667em\lower.7ex\hbox{E}\kern-.125emX}}

\newcommand{\qbinom}[2]{\genfrac{[}{]}{0pt}{}{\,#1\,}{\,#2\,}_{\!q}}     
    
\begin{document}

\title{Decoding Probability Analysis of Network-Coded Data Collection and Delivery by Relay Drones}

\author{\IEEEauthorblockN{Ioannis~Chatzigeorgiou and Elena~Manole}
\IEEEauthorblockA{School of Computing and Communications\\
Lancaster University\\Lancaster LA1 4WA, United Kingdom\\
Email: \{i.chatzigeorgiou, e.manole\}@lancaster.ac.uk}}

\maketitle

\begin{abstract}
Relay drones in delay-tolerant applications are dispatched to remote locations in order to gather data transmitted by a source node. Collected data are stored on the drones and delivered to one or multiple bases. This paper considers two schemes for broadcasting data to drones when feedback channels are not available: a \textit{data carousel} and \textit{systematic random linear network coding} (RLNC). We propose a theoretical framework for the calculation of the probability that a base will fully or partially recover the transmitted data and the probability that all involved bases will successfully obtain the data, when the~bases are either isolated or interconnected. Theoretical results are validated through simulations. Design considerations are also discussed, including the relationship among the field size used by RLNC, the number of relay drones and the requirement for full data recovery or the retrieval of at least part of the data.
\end{abstract}

\vspace{4pt}
\begin{IEEEkeywords}
Unmanned aerial vehicles, random linear network coding, data carousel, fading channel, mission success.
\end{IEEEkeywords}

\section{Introduction}
\label{sec:intro}

Although unmanned aerial vehicles, commonly referred to as \textit{drones}, were initially considered for military applications, such as surveillance, reconnaissance and rescue operations, their use has been extended to commercial and civil applications, including environmental monitoring, maritime safety, remote sensing and communications. Drones serving as flying base stations or communication relays have the potential to enhance wireless networks and enable future services~\cite{Zeng2016,Mozaffari2019}. 

The focus of this paper is on \textit{relay drones}, which provide on-demand wireless connectivity between remote network nodes when the line-of-sight channel is blocked, e.g. due to physical obstacles, or is compromised, e.g. due to intentional jamming or unintentional interference. In essence, drones offer another way to ``put antennas in the sky'' and lower the demand for limited satellite resources~\cite{Carr2009}. In delay-tolerant applications, relay drones are not always required to establish an end-to-end communication path between two nodes. They can be dispatched from multiple bases to a field site with the objective of collecting information transmitted by an on-site data-gathering source node. For example, the source node could be a wireless sensor forwarding measurements for precision agriculture, or a reconnaissance team sending critical data for emergency or military operations. Collected information is stored on the drones, which carry and deliver it to their designated bases for further processing.

This work considers the broadcasting of information by a source node to clusters of drones in the absence of feedback. The sequential transmission and periodical repetition of data packets, known as a \textit{data carousel}~\cite{Bulut2013}, is compared  to \textit{random linear network coding} (RLNC) at the source node \cite{Byers1998,Ho-RNC}, in terms of the probability that a particular base or all of the bases will fully or partially recover the broadcast information. The motivation for this paper is to characterize the performance of a data carousel, bring together recent theoretical advances in RLNC, e.g. \cite{Jones15,Claridge2017,Tsimbalo2018} and develop a framework tailored for broadcast communication aided by relay drones that collect, carry and deliver received packets. The theoretical framework can then be used to quantify performance trade-offs in the two transmission schemes.

The remainder of this paper has been organized as follows: Section~\ref{sec:system} presents the system model, describes the system configurations and proposes performance metrics. Sections~\ref{sec:isolated} and \ref{sec:interconnected} analyze the system configurations and present analytic expressions for the performance metrics. The theoretical framework is validated through simulations in Section~\ref{sec:results} and performance comparisons are discussed. Key findings and conclusions are summarized in Section~\ref{sec:conclusions}.

\section{System Model and Problem Formulation}
\label{sec:system}

\begin{figure}[t]
\centering
\includegraphics[width=0.85\columnwidth]{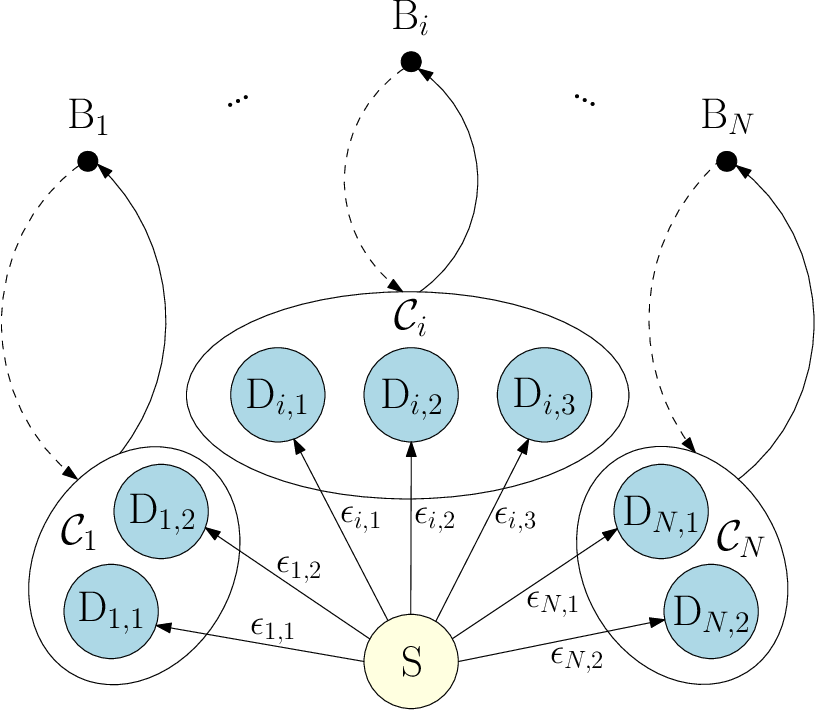}
\caption{Depiction of a source node $\mathrm{S}$ transmitting packets to drone clusters $\mathcal{C}_1,\dots,\mathcal{C}_N$ over a broadcast erasure channel. At the end of the transmission, each cluster $\mathcal{C}_i$ returns to base $\mathrm{B}_i$ to offload received packets.}
\label{fig:system_model}
\end{figure}

We consider a source node $\mathrm{S}$ attempting to deliver a message to $N$ clusters $\mathcal{C}_1,\dots,\mathcal{C}_N$ of drones, as shown in Fig.~\ref{fig:system_model}. The number of drones in cluster $\mathcal{C}_i$ is given by $L_i=|\mathcal{C}_i|$, for $i=1,\dots,N$, while the total number of receiving drones is $L=\sum_i L_i$. The $j$-th drone in cluster $\mathcal{C}_i$ is denoted by $\mathrm{D}_{i,j}$. Each drone in $\mathcal{C}_i$ transports and relays information about the source message to base $\mathrm{B}_i$. The $N$ bases can be either \textit{isolated} or \textit{interconnected}. In the former case, base $\mathrm{B}_i$ relies entirely on the information relayed by cluster $\mathcal{C}_i$ to reconstruct the source message. In the latter case, each base has access to the information received by the other $N-1$ bases.

Before transmission, the message at node $\mathrm{S}$ is segmented into $k$ source packets, $u_1,\ldots,u_k$. Node $\mathrm{S}$ subsequently broadcasts a sequence of $n_\mathrm{T}$ packets, $x_1,\ldots,x_{n_\mathrm{T}}$. The broadcast sequence is composed of the $k$ source packets followed by $n_\mathrm{T}-k$ packets, that is, $x_n=u_n$ for $n=1,\dots,k$, while the content of $x_n$ for $n=k+1,\dots,n_\mathrm{T}$ depends on the employed transmission method. In this paper, we assume that feedback channels between the drones and node $\mathrm{S}$ are not available. For instance, this could be the case when the adopted communication protocol does not support the reliable broadcasting of data, as in IEEE 802.11. Even when feedback mechanisms are supported, the feedback links of the power-constrained drones could be prone to high rates of packet loss due to interference or intentional jamming. A \textit{data carousel} \cite{Bulut2013} is a conventional way of improving data reliability in unidirectional broadcast environments by repeatedly transmitting the source packets in a cyclic fashion. The relationship between the source packets and the transmitted packets can be expressed as:
\begin{equation}
\label{eq.Carousel_Packet}
x_{n} = u_{\,((n-1)\bmod k)\,+\,1}\;\;\mathrm{for}\;\; n=k+1,\dots,n_\mathrm{T},
\end{equation}
where $\bmod$ denotes the modulo operator. On the other hand, when \textit{systematic RLNC} \cite{Shrader09,Jones15} is employed, the $k$ source packets are followed by $n_\mathrm{T}-k$ coded packets, which are random linear combinations of the $k$ source packets. A coded packet $x_n$ can be obtained as follows:
\begin{equation}
\label{eq.SysRLNC_Packet}
x_{n} = \sum_{b=1}^{k} c_{n,b}\,u_b\;\;\mathrm{for}\;\; n=k+1,\dots,n_\mathrm{T},
\end{equation}
where each coefficient $c_{n,b}$ is chosen uniformly at random from a finite field of $q$ elements, denoted by $\mathrm{GF}(q)$, for $q$ a prime power.

The $n_\mathrm{T}$ packets are transmitted over a broadcast erasure channel, where $\epsilon_{i,j}$ is the packet erasure probability between the source node $\mathrm{S}$ and drone $\mathrm{D}_{i,j}$, as depicted in Fig.~\ref{fig:system_model}. The packet erasure probability captures the channel conditions as well as the modulation and coding scheme (MCS) used by the transmitting source node and the receiving drones. In cases where the Nakagami-$m$ fading model can accurately describe the channel between node $\mathrm{S}$ and drone $\mathrm{D}_{i,j}$, e.g. as in \cite{Yanmaz2013}, the packet erasure probability can be expressed in analytic form as a function of MCS characteristics and channel parameters~\cite{Xi2011}. It can also be approximated by:
\begin{equation}
\label{eq.PER_Nakagami}
\epsilon_{i,j} \approx \left(\frac{m}{\overline{\gamma}_{i,j}}\right)^{\!m} \frac{w_m}{\mathrm{\Gamma}(m)},
\end{equation}
where $m\geq 0.5$ is the shape factor of the Nakagami distribution, $\Gamma(\cdot)$ is the Gamma function, $\overline{\gamma}_{i,j}$ is the average signal-to-noise ratio (SNR) of the channel between node $\mathrm{S}$ and drone $\mathrm{D}_{i,j}$, and $w_m$ is an SNR threshold, which is specific to the employed MCS and can be computed using \cite[eq.~(13)]{Xi2011}.

After the source node has transmitted the $n_\mathrm{T}$ packets, the drones of cluster $\mathcal{C}_i$ do not attempt to reconstruct the source message but return to base $\mathrm{B}_i$ to deliver their collected packets. A base will reconstruct the source message, if the $k$ source packets can be recovered from the packets delivered by the respective cluster of drones. The mission is deemed a success if all of the bases reconstruct the source message.

The following two sections present expressions for the probability of a base decoding a source message, either fully or partially, and the probability of mission success. The analysis considers a source node using either a data carousel or systematic RLNC to broadcast packets to clusters of drones, which transport and deliver received packets to isolated or interconnected bases.

\section{Analysis for Isolated Bases}
\label{sec:isolated}

Recall that a packet transmitted by the source node $\mathrm{S}$ will be received by drone $\mathrm{D}_{i,j}$ with probability $1-\epsilon_{i,j}$. Base $\mathrm{B}_i$ will eventually acquire all packets that have been collected by drones $\mathrm{D}_{i,1},\dots,\mathrm{D}_{i,L_i}$ in cluster $\mathcal{C}_i$. The probability that base $\mathrm{B}_i$ will not acquire a packet is the probability of that packet being `erased' (i.e. not received) by all of the drones in cluster $\mathcal{C}_i$. We can thus write the \textit{equivalent packet erasure probability} $\epsilon_i$ experienced by base $\mathrm{B}_i$ as:
\begin{equation}
\label{eq.equiv_erasure_prob}
\epsilon_i = \prod_{j=1}^{L_i}\epsilon_{i,j}.
\end{equation}
Similarly, the probability that base $\mathrm{B}_i$ will obtain a particular packet is the probability that at least one of the drones in cluster $\mathcal{C}_i$ will receive it, given by $(1-\epsilon_i)=1-\prod_{j=1}^{L_i}\epsilon_{i,j}$.

\subsection{Transmission based on a Data Carousel}
\label{subsec:isolated_DC}

Let $\lambda$ and $\rho$ be the quotient and remainder, respectively, of the division of the number of transmitted packets $n_\mathrm{T}$ by the number of source packets $k$, that is, $n_\mathrm{T}=\lambda k + \rho$~where \mbox{$\lambda=\lfloor n_\mathrm{T}/k\rfloor$} and $\rho=n_\mathrm{T}\bmod k$. The sequence of $n_\mathrm{T}$ transmitted packets will thus consist of $\lambda$ copies of the $k$ source packets and one additional copy of the first $\rho$ source packets. In total, $\lambda+1$ copies of the first $\rho$ source packets and $\lambda$ copies of the remaining $k-\rho$ source packets will be broadcast.

The probability of base $\mathrm{B}_i$ reassembling the source message can be expressed as the product of two probabilities: (i) the probability that at least one copy, out of $\lambda+1$ copies, of each of the first $\rho$ source packets will be obtained, and (ii) the probability that at least one copy, out of  $\lambda$ copies, of each of the remaining $k-\rho$ source packets will be retrieved. Therefore, we can write:
\begin{equation}
\label{eq.prob_carousel_cluster_full}
P_\mathrm{dc}(\epsilon_i) = (1-\epsilon_i^{\lambda+1})^{\rho}\:(1-\epsilon_i^{\lambda})^{k-\rho},
\end{equation}
where `data carousel' has been abbreviated to `dc'.

Given that feedback links are not available, the reliability of the broadcast channel does not affect the order with which packets are being broadcast by the source node. When the $N$ bases are isolated (`iso' for brevity), they will all reconstruct the source message, if and only if each one of them obtains the $k$ source packets. Consequently, the probability of mission success can be expressed as:
\begin{equation}
\label{eq.prob_carousel_mission_full}
P_{\mathrm{iso}\text{-}\mathrm{dc}}(\boldsymbol{\epsilon})=\prod_{i=1}^{N}P_\mathrm{dc}(\epsilon_{i}),
\end{equation}
where $\boldsymbol{\epsilon}=(\epsilon_{1},\ldots,\epsilon_{N})$ is a vector containing the equivalent erasure probability experienced by each base.

\subsection{Transmission based on Systematic RLNC}
\label{subsec:isolated_SR}

Base $\mathrm{B}_i$ will be able to reconstruct the $k$ source packets of the message, if the mix of source and coded packets that have been collected by the drones in cluster $\mathcal{C}_i$ contain $k$ linearly independent packets. Let $P_\mathrm{sr}(k, n)$ denote the probability that a base will recover the $k$ source packets when the drones of a cluster have deposited $n$ received packets at that base, where $k\leq n\leq n_\mathrm{T}$. Index `sr' in $P_\mathrm{sr}(k, n)$ is used as an abbreviation for `systematic RLNC'. An expression for $P_\mathrm{sr}(k, n)$ can be obtained from \cite{Shrader09,Jones15} and can take the form:
\begin{equation}
\label{eq.cond_sys_full_rank}
P_\mathrm{sr}(k,n)=%
\sum_{h=h_\mathrm{low}}^{k}\!\!\!
\resizebox{0.11\textwidth}{!}{$
\frac{\displaystyle\binom{k}{h}%
\binom{n_\mathrm{T}-k}{n-h}}%
{\displaystyle\binom{n_\mathrm{T}}{n}}
$}
\prod_{w=0}^{k-h-1}\left(1-q^{-n+h+w}\right),%
\end{equation}
where $h_\mathrm{low}=\max{(0,\,n-n_\mathrm{T}+k)}$. Note that \eqref{eq.cond_sys_full_rank} quantifies the probability that a base will obtain $h$ source (hence, linearly independent) packets and $n-h$ coded packets, among which $k-h$ will be linearly independent, for all valid values of $h$. If a particular cluster $\mathcal{C}_{i}$ is considered, which experiences packet erasures with equivalent probability $\epsilon_{i}$ as defined in \eqref{eq.equiv_erasure_prob}, the average probability of base $\mathrm{B}_i$ recovering the \textit{full} message is given by \cite{Jones15}:
\begin{equation}
\label{eq.prob_BEC}
P_\mathrm{sr}(\epsilon_{i})=\sum_{n=k}^{n_\mathrm{T}}\binom{n_\mathrm{T}}{n}\left(1-\epsilon_{i}\right)^{n}\epsilon_{i}^{n_\mathrm{T}-n}\,P_\mathrm{sr}(k,n).
\end{equation}

For RLNC over fields of large size, base~$\mathrm{B}_i$ is highly likely to recover the $k$ source packets as soon as $k$ different packets are deposited at $\mathrm{B}_i$. That is, $P_\mathrm{sr}(k,n)\rightarrow 1$ for $q\rightarrow\infty$~\cite{Eryilmaz2006, Lucani2009}. As a result, $P_\mathrm{sr}(\epsilon_{i})$ in \eqref{eq.prob_BEC} reduces to a complementary binomial cumulative distribution function, while the network-coded broadcast flow behaves like a collection of independent network-coded unicast flows. The probability of delivering the full message to the $N$ isolated bases, which is the probability of mission success, can thus be approximated by:
\begin{equation}
\label{eq.prob_del_approx}
P_{\mathrm{iso}\text{-}\mathrm{sr}}(\boldsymbol{\epsilon})\approx\prod_{i=1}^{N}P_\mathrm{sr}(\epsilon_{i}),\;\text{for}\,\,q\rightarrow\infty\,\,\text{and}\,\,P_\mathrm{sr}(k,n)\rightarrow 1,
\end{equation}
where $\boldsymbol{\epsilon}=(\epsilon_{1},\ldots,\epsilon_{N})$. Approximation~\eqref{eq.prob_del_approx} can be extended to non-systematic RLNC but can be loose, especially when finite fields of small size are used, as explained in \cite{Tsimbalo2018}. When systematic RLNC is employed, approximation \eqref{eq.prob_del_approx} becomes a lower bound, that is \cite[Theorem 2]{Tsimbalo2018}:
\begin{equation}
\label{eq.prob_del_bound}
P_{\mathrm{iso}\text{-}\mathrm{sr}}(\boldsymbol{\epsilon})\geq\prod_{i=1}^{N}P_\mathrm{sr}(\epsilon_{i}),\;\text{for}\;\textit{any}\;\text{valid}\,\,q.
\end{equation}
The bound becomes tighter for increasing values of $q$ and $k$, and decreasing values of $N$ and $\epsilon_{i}$. As reported in \cite{Tsimbalo2018}, the mean squared error (MSE) between the exact probability value -- computed via simulations -- and the lower bound is $9\cdot 10^{-5}$ for $q=2$, $k=20$, $N=6$ and $\boldsymbol{\epsilon}=(0.1,\ldots,0.1)$. The MSE drops to $5\cdot 10^{-6}$ when the packet erasure probability reduces to $\boldsymbol{\epsilon}=(0.01,\ldots,0.01)$. 

\section{Analysis for Interconnected Bases}
\label{sec:interconnected}

The probability of mission success can greatly improve if the $N$ bases are interconnected, that is, each base can share all acquired packets with every other base at the expense of increased latency, which should not exceed a target value set by the system requirements. Even if the full source message may be reconstructed by the interconnected bases within an acceptable time frame, each base could have the additional requirement of recovering at least part of the message as soon as the drones of the respective cluster offload their received packets. In this section, in addition to the probability of mission success, we present expressions for the probability of a base recovering at least $\mu$ of the $k$ source packets when it has obtained $n$ of the $n_T$ transmitted packets, where $\mu\leq k$ and $\mu\leq n\leq n_\mathrm{T}$.

\subsection{Transmission based on a Data Carousel}
\label{subsec:interconnected_DC}

Let $Y_1$ and $Y_2$ be discrete random variables; $Y_1$ represents the number of source packets from the set $\{u_1,\ldots,u_\rho\}$ that have been retrieved by a base, whereas $Y_2$ represents the number of source packets from the set $\{u_{\rho+1},\ldots,u_k\}$ that have been acquired by the same base. The joint probability mass function of $Y_1$ and $Y_2$, denoted by $p_{Y_1,Y_2}(y_1,y_2)$, is the probability that a base will recover \textit{exactly} $Y_1=y_1$ and $Y_2=y_2$ source packets. Recall from Section~\ref{subsec:isolated_DC} that the number of packets transmitted by the source node can be expressed as $n_\mathrm{T}=\lambda k + \rho$. 

Base $\mathrm{B}_i$ will acquire any $y_1$ of the first $\rho$ source packets, i.e. $u_1,\ldots,u_\rho$, if at least one copy, out of $\lambda+1$ copies, of each of these $y_1$ source packets reaches the base, while all of the copies of the remaining $\rho-y_1$ source packets are erased. The same base will obtain any $y_2$ of the last $k-\rho$ source packets, if at least one copy, out of $\lambda$ copies, of each of these $y_2$ source packets is offloaded at the base, but no copies of the remaining $(k-\rho)-y_2$ source packets are received. Therefore, the joint probability mass function can be written as:
\begin{align}
\label{eq.prob_dc_cluster_part_v1}
p_{Y_1,Y_2}(y_1,\,y_2) = \:&%
\binom{\rho}{y_1}
(1-\epsilon_i^{\lambda+1})^{y_1}\:\epsilon_i^{(\lambda+1)(\rho-y_1)}\notag\\
&\cdot 
\binom{k-\rho}{y_2}
(1-\epsilon_i^{\lambda})^{y_2}\:\epsilon_i^{\lambda\,[\,(k-\rho)-y_2\,]},
\end{align}
which reduces to:
\begin{align}
&\!\!p_{Y_1,Y_2}(y_1,\,y_2) = \notag \\
&\!\!\!=\!%
\resizebox{0.11\textwidth}{!}{$
\displaystyle\binom{\rho}{y_1}\!\binom{k-\rho}{y_2}
$}
\resizebox{0.31\textwidth}{!}{$
(1-\epsilon_i^{\lambda+1})^{y_1}(1-\epsilon_i^{\lambda})^{y_2}\:\epsilon_i^{\lambda(k-y_1-y_2)+\rho-y_1}
$}\!.
\end{align}
The probability of base $\mathrm{B}_i$ recovering \textit{at least} $\mu$ of the $k$ source packets can be expressed in terms of the joint probability mass function as follows:
\begin{equation}
\label{eq.prob_carousel_cluster_part}
P_\mathrm{dc}^{\:(\mu/k)}(\epsilon_i)\:=
\:\sum_{y=\mu}^{k}\:\sum_{y_1=y_\mathrm{low}}^{y_\mathrm{hi}}p_{Y_1,Y_2}(y_1,\,y\!-\!y_1),
\end{equation}
where $y_\mathrm{low}=\max(0,y-k+\rho)$ and $y_\mathrm{hi}=\min(y,\rho)$. The inner sum in \eqref{eq.prob_carousel_cluster_part} computes the probability that base $\mathrm{B}_i$ will recover \textit{exactly} $y$ of the $k$ source packets, for all valid values of $y_1$ and $y_2$ that give $y_1+y_2=y$. Note that $y_2$ in $p_{Y_1,Y_2}(y_1,y_2)$ has been written as $y_2=y-y_1$. The outer sum aggregates these probabilities for $y$ ranging from $\mu$ to $k$. For $\mu=k$, we obtain $y=k$, $y_1=\rho$ and equation \eqref{eq.prob_carousel_cluster_part} collapses to \eqref{eq.prob_carousel_cluster_full}, which gives the probability of base $\mathrm{B}_i$ acquiring all of the source packets.

To compute the probability of mission success, the group of $N$ interconnected bases can be viewed as a single base for packet offloading, while the union of the $N$ clusters can be treated as a superset of $L$ drones. The system model reduces to a source node broadcasting packets to a single cluster of $L$ drones, which offload received packets at a single base experiencing equivalent erasure probability:
\begin{equation}
\label{eq.equiv_erasure_prob_interconnected}
\epsilon = \prod_{i=1}^{L}\epsilon_{i}.
\end{equation}
Hence, the probability that the $N$ interconnected bases will obtain the $k$ source packets and reconstruct the source message can be calculated using:
\begin{equation}
\label{eq.prob_carousel_mission_part}
P_{\mathrm{int}\text{-}\mathrm{dc}}(\epsilon)=P_\mathrm{dc}(\epsilon),
\end{equation}
where `int' is short for `interconnected' and $P_\mathrm{dc}(\cdot)$ is given by \eqref{eq.prob_carousel_cluster_full}.

\subsection{Transmission based on Systematic RLNC}
\label{subsec:interconnected_SR}

Literature on RLNC is primarily concerned with the probability of a receiver recovering all of the $k$ source packets and, hence, reassembling the full transmitted message. The probability of obtaining a fraction of the message was investigated in \cite{Claridge2017} and can be extended to the system model under consideration. In particular, the probability that a base will recover at least $\mu$ source packets from $n$ packets offloaded by a cluster of drones, can be obtained from \cite[Proposition 2]{Claridge2017}:
\begin{align}
\label{eq.cond_sys_rank_def}
&P_\mathrm{sr}(\mu,k,n)=%
\frac{1}{\binom{n_\mathrm{T}}{n}}
\cdot\nonumber%
\\
&\resizebox{0.442\textwidth}{!}{$
\cdot\!\!\!\displaystyle\sum_{r=\mu}^{\min(n,k)}\!\!\!%
\sum_{h=h_\mathrm{low}}^{r}\!\!\Biggl(%
\binom{k}{h}\binom{n_\mathrm{T}-k}{n-h}%
q^{-(n-h)(k-r)}%
\!\!\prod_{w=0}^{r-h-1}%
(1-q^{-n+h+w})%
\cdot$}\nonumber%
\\
&\resizebox{0.436\textwidth}{!}{$
\cdot\!\!\displaystyle\sum_{b=b_\mathrm{low}}^{r-h}%
\!\!\!\binom{k-h}{b}%
\!\sum_{\ell=0}^{k-h-b}%
\left(-1\right)^{\ell}%
\binom{k-h-b}{\ell}%
\qbinom{k-h-b-\ell}{r-h-b-\ell}%
\Biggr)%
$}
\end{align}
where $h_\mathrm{low}=\max{(0,n-n_\mathrm{T}+k)}$, $b_\mathrm{low}=\max(0,\mu-h)$ and $\qbinom{b}{\ell}$ is the \textit{Gaussian binomial coefficient} defined by \cite[p. 125]{Cameron1994}:
\begin{equation}
\label{eq.def_GBCoeff}
\setlength{\nulldelimiterspace}{0pt}
\qbinom{b}{\ell}=\left\{%
\begin{array}{cc}
\displaystyle\prod_{i=0}^{\ell-1}\frac{(q^b-q^i)}{(q^\ell-q^i)},&\mathrm{for}\;\; \ell\leq b,\\
0,&\mathrm{for}\;\; \ell>b.%
\end{array}\right.
\end{equation}
Expression~\eqref{eq.cond_sys_rank_def} enumerates all combinations of $r\geq\mu$ linearly independent received packets, which consist of $h$ source packets and $r-h$ coded packets. The coded packets can be decoded into at least $\mu-h$ source packets. If we substitute \eqref{eq.cond_sys_rank_def} into \eqref{eq.prob_BEC}, the probability that base $\mathrm{B}_i$ will recover at least $\mu$ of the $k$ source packets and will fully (for $\mu=k$) or partially (for $0<\mu<k$) reconstruct the source message, is given by:
\begin{equation}
\label{eq.prob_BEC_part}
P_\mathrm{sr}^{\:(\mu/k)}(\epsilon_i)=\sum_{n=k}^{n_\mathrm{T}}\binom{n_\mathrm{T}}{n}\left(1-\epsilon_i\right)^{n}\epsilon_i^{n_\mathrm{T}-n}\,P_\mathrm{sr}(\mu,k,n).
\end{equation}
Note that for $\mu=k$, then $P_\mathrm{sr}(\mu,k,n)$ in \eqref{eq.cond_sys_rank_def} reduces to $P_\mathrm{sr}(k,n)$ in \eqref{eq.cond_sys_full_rank}. Therefore, \eqref{eq.cond_sys_rank_def} and \eqref{eq.prob_BEC_part} can be viewed as generalisations of \eqref{eq.cond_sys_full_rank} and \eqref{eq.prob_BEC}, respectively.

Similar to the data carousel, if the $N$ bases are interconnected, the system reduces to a source node broadcasting packets to a single cluster of $L$ drones, which receive, transport and deliver packets to a single base. Therefore, the probability that the $N$ interconnected bases will fully reconstruct the source message and the mission will be successfully completed is simply:
\begin{equation}
\label{eq.prob_del_interconnected}
P_{\mathrm{int}\text{-}\mathrm{sr}}(\epsilon)=P_\mathrm{sr}(\epsilon),
\end{equation}
where $P_\mathrm{sr}(\cdot)$ is given by \eqref{eq.prob_BEC} and $\epsilon$ is the equivalent average packet erasure probability experienced by the union of all clusters, defined in \eqref{eq.equiv_erasure_prob_interconnected}.

\section{Results and Discussion}
\label{sec:results}

\begin{figure*}[!t]
\centerline{\subfloat[Isolated bases]{\includegraphics[width=8.65cm]{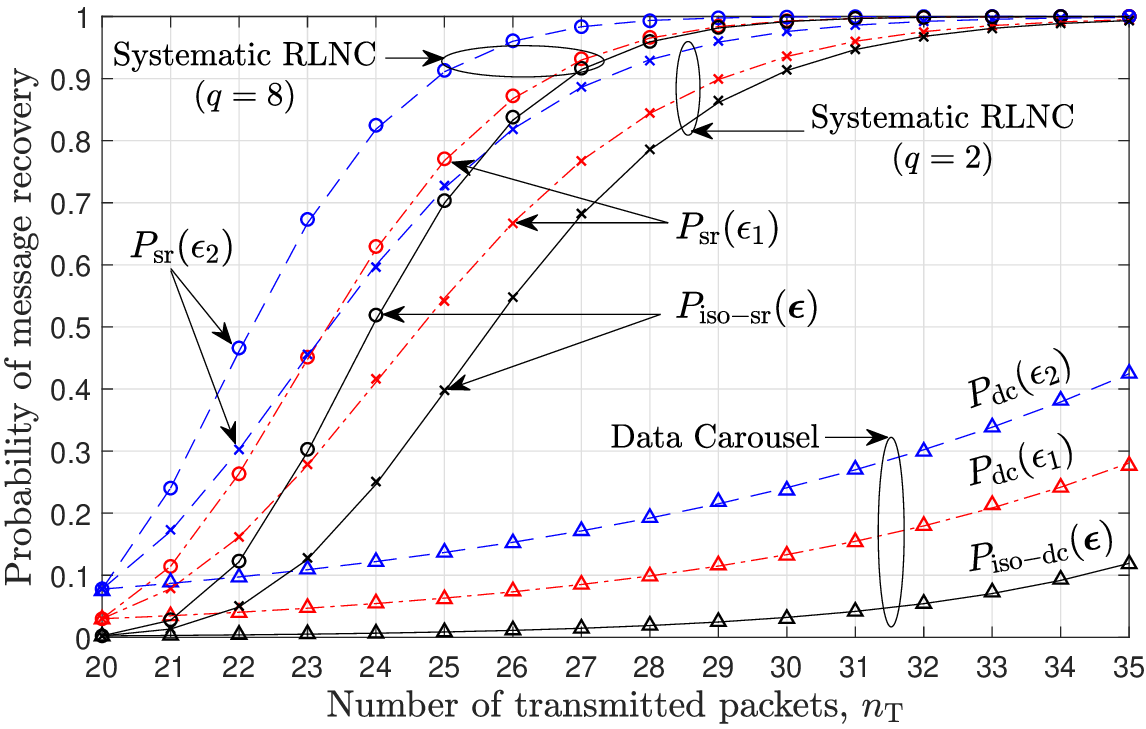}
\label{fig_isolated}}
\hfil
\subfloat[Interconnected bases]{\includegraphics[width=8.65cm]{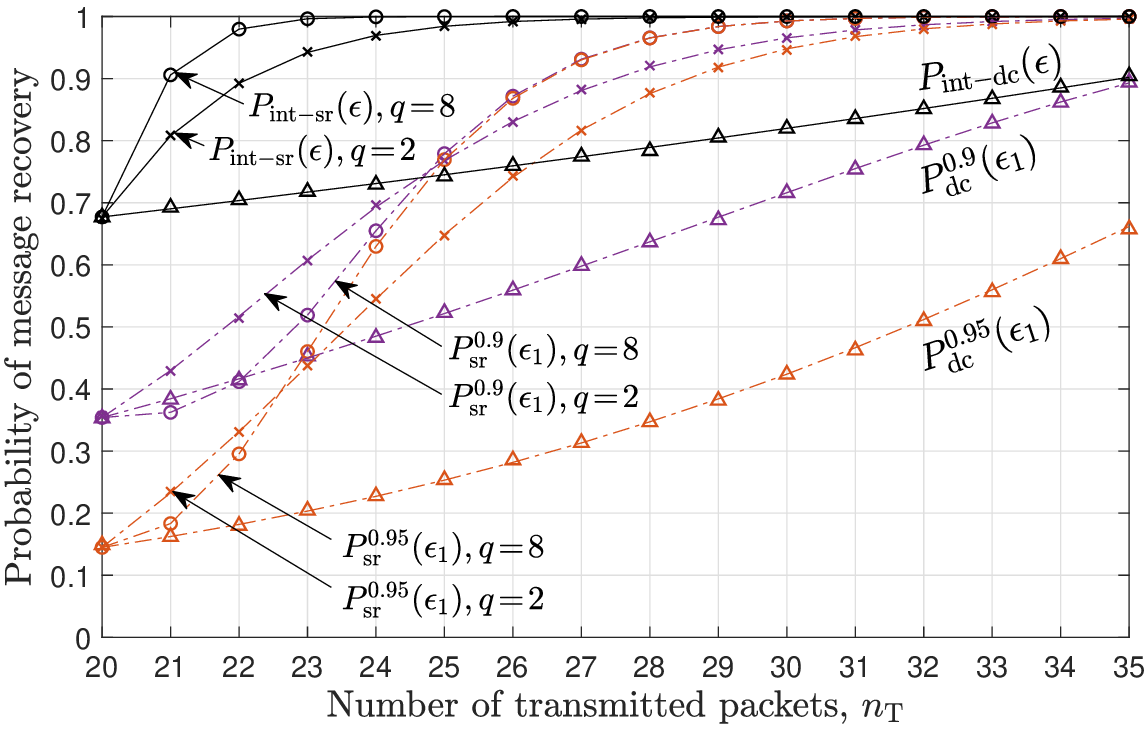}
\label{fig_interconnected}}}
\caption{Comparison of theoretical values (curves) and simulation results (markers on curves) for the probability of mission success and the probability of a base, either $\mathrm{B}_1$ or $\mathrm{B}_2$, recovering a source message of $k=20$ packets for an increasing value of transmitted packets ($n_\mathrm{T}$). The probability of mission success for a data carousel or systematic RLNC is shown in (a) for isolated bases, given by $P_{\mathrm{iso}\text{-}\mathrm{dc}}(\boldsymbol{\epsilon})$ and $P_{\mathrm{iso}\text{-}\mathrm{sr}}(\boldsymbol{\epsilon})$, respectively, and (b) for interconnected bases, given by $P_{\mathrm{int}\text{-}\mathrm{dc}}(\epsilon)$ and $P_{\mathrm{int}\text{-}\mathrm{sr}}(\epsilon)$, respectively. Curves labeled as $P_{z}(\epsilon_i)$ or $P^{\alpha}_{z}(\epsilon_i)$ for $z\in\{\mathrm{dc},\mathrm{sr}\}$ depict the probability that base $\mathrm{B}_i$, for $i\in\{1,2\}$, will recover all of the source packets or at least $\alpha k$ of the source packets, irrespective of whether the two bases are isolated or interconnected.}
\label{fig_validation}
\end{figure*}

In order to demonstrate the exactness of the theoretical framework, the model of a system consisting of one source node, two clusters of drones and two bases was developed in MATLAB. Cluster $\mathcal{C}_1$ contained $L_1=3$ drones, while cluster $\mathcal{C}_2$ comprised $L_2=2$ drones. The packet erasure~probabilities at the five relay drones were set to $\epsilon_{1,1}=0.45$, $\epsilon_{1,2}=0.55$, $\epsilon_{1,3}=0.65$, $\epsilon_{2,1}=0.3$ and $\epsilon_{2,2}=0.4$. A message of $k=20$ packets was broadcast by the source node using a data carousel or systematic RLNC over $\mathrm{GF}(q)$, for $q\in\{2,8\}$. For each transmission method, 50000 experiments involving the transmission of $n_\mathrm{T}$ packets by the source node to the two bases, via the two clusters of drones, were carried out for $n_\mathrm{T}=20,\dots,35$. Successful missions and instances of a base recovering at least some or all of the source packets were counted and averaged over all experiments. Plots of the measured probabilities, obtained through simulations, and the computed probabilities, obtained from the analytic expressions in Section~\ref{sec:isolated} and Section~\ref{sec:interconnected}, are presented in Fig.~\ref{fig_validation}.

As can be seen in Fig.~\ref{fig_isolated}, when the two bases are isolated and a data carousel is used for broadcasting, an increase in the number of transmitted packets causes only a small improvement in the probability of mission success, denoted by $P_{\mathrm{iso}\text{-}\mathrm{dc}}(\boldsymbol{\epsilon})$ and computed using \eqref{eq.prob_carousel_mission_full}. As expected, base $\mathrm{B}_2$ is more likely to reconstruct the source message than base~$\mathrm{B}_1$, as the relationship between the equivalent packet erasure probabilities $\epsilon_1$ and $\epsilon_2$, obtained from \eqref{eq.equiv_erasure_prob}, is $\epsilon_2<\epsilon_1$, therefore  $P_\mathrm{dc}(\epsilon_2)>P_\mathrm{dc}(\epsilon_1)$. For systematic RLNC over $\mathrm{GF}(2)$, the value of $n_\mathrm{T}$ has a notable impact on the probability of mission success, i.e. $P_{\mathrm{iso}\text{-}\mathrm{sr}}(\boldsymbol{\epsilon})$, which is bounded by \eqref{eq.prob_del_bound}. As is the case with RLNC, this impact becomes more pronounced when $\mathrm{GF}(8)$ is used. Similar observations can be made for the probability that base $\mathrm{B}_i$, for $i\in\{1,2\}$, will retrieve all of the source packets, given by $P_\mathrm{sr}(\epsilon_i)$ in \eqref{eq.prob_BEC}. In all cases, simulation results are in agreement with analytic expressions.

If the two bases are interconnected, the probability of mission success improves significantly, as shown in Fig.~\ref{fig_interconnected}. In the case of a data carousel, $P_{\mathrm{int}\text{-}\mathrm{dc}}(\epsilon)$ is close to $90\%$ for $n_\mathrm{T}=35$. Systematic RLNC requires fewer packet transmissions to achieve $P_{\mathrm{int}\text{-}\mathrm{sr}}(\epsilon)\approx 90\%$, i.e. $n_\mathrm{T}=22$ for $q=2$, while $n_\mathrm{T}=21$ for $q=8$. Recall that $P_{\mathrm{int}\text{-}\mathrm{dc}}(\epsilon)$ and $P_{\mathrm{int}\text{-}\mathrm{sr}}(\epsilon)$ can be computed using \eqref{eq.prob_carousel_mission_part} and \eqref{eq.prob_del_interconnected}, respectively. Fig.~\ref{fig_interconnected} also depicts the probability that base $\mathrm{B}_1$ will recover at least $\mu=18$ or $\mu=19$ of the $k=20$ source packets, thus $\mu/k$ is set to $0.9$ and $0.95$, respectively, in \eqref{eq.prob_carousel_cluster_part} and \eqref{eq.prob_BEC_part}. Both $P^{\mu/k}_{\mathrm{dc}}(\epsilon_1)$ and $P^{\mu/k}_{\mathrm{sr}}(\epsilon_1)$ are computed before any information is exchanged between bases. Therefore, the respective curves in Fig.~\ref{fig_interconnected} would remain the same if the bases were isolated. A noteworthy observation is that RNLC over large finite fields yields a high probability of full message recovery but offers a low probability of partial message recovery for a small number of transmitted packets. For example, base $\mathrm{B}_1$ stands a higher chance of recovering at least $18$ of the $20$ source packets $(\mu/k=0.9)$ if $\mathrm{GF}(2)$ is chosen over $\mathrm{GF}(8)$ when $21\leq n_\mathrm{T}\leq 24$, as shown in Fig.~\ref{fig_interconnected}. 

\begin{figure}[t]
\centering
\includegraphics[width=8.65cm]{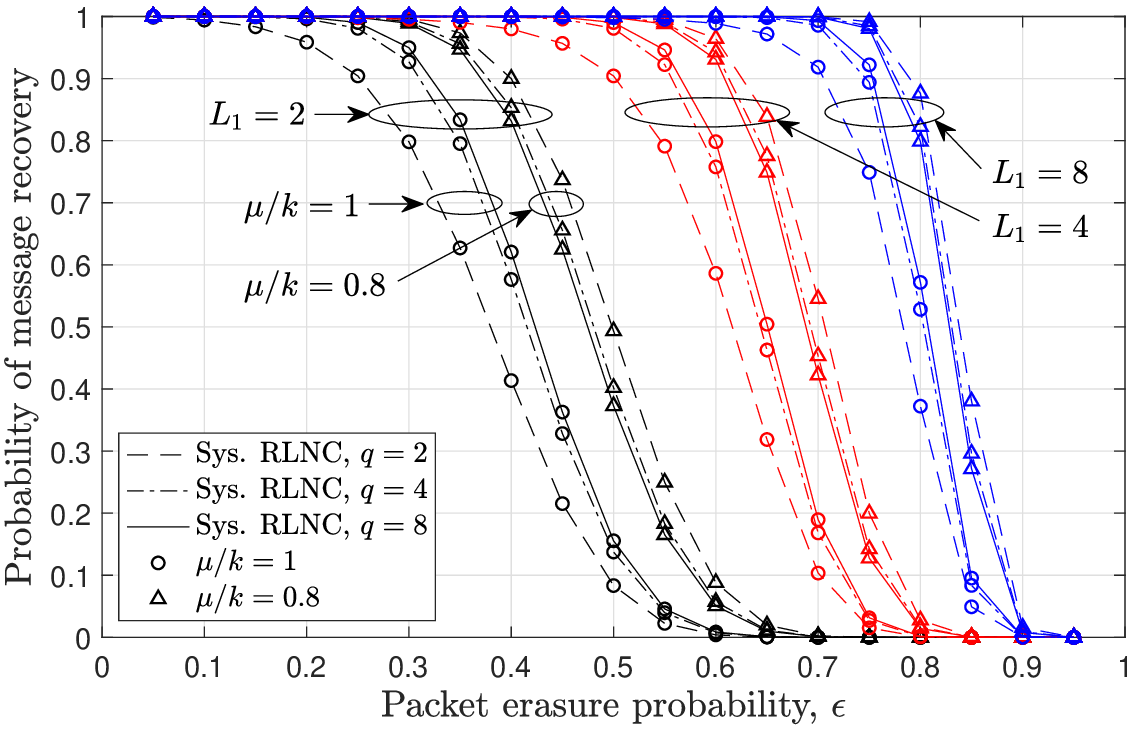}
\caption{Probability of base $\mathrm{B}_1$ partially $(\mu/k=0.8)$ or fully \mbox{$(\mu/k=1)$} \mbox{recovering} a message of $k\!=\!30$ source packets. Systematic RLNC~over~$\mathrm{GF}(q)$ is used to broadcast $n_\mathrm{T}=36$ packets to $L_1$ drones, where $q\in\{2,4,8\}$ and $L_1\in\{2,4,8\}$.}
\label{fig:drones_vs_erasure}
\end{figure}

Having validated the proposed theoretical framework, we illustrate with examples how it can contribute to system~design. In order to further investigate the observation made in Fig.~\ref{fig_interconnected}, we concentrate on systematic RLNC over $\mathrm{GF}(q)$ and look into the impact of the field size $q$ on the probability that a base will partially or fully recover the source message. Let us focus on a base, e.g. $\mathrm{B}_1$, which collects the packets received by a cluster of $L_1$ drones, where $L_1\in\{2,4,8\}$. For simplicity, we have set $\epsilon_{1,1}=\ldots=\epsilon_{1,L_1}=\epsilon$, where the packet erasure probability $\epsilon$ takes values in the range $[0.05,0.95]$. Fig.~\ref{fig:drones_vs_erasure} depicts the probability that base $\mathrm{B}_1$ will recover at least $\mu=24$ of the $k=30$ source packets $(\mu/k=0.8)$ or all of the source packets $(\mu/k=1)$, when $q\in\{2,4,8\}$ and $n_\mathrm{T}=36$. The figure shows that, for a given $L_1$, the probability of full message recovery improves for an increasing value of $q$, but the trend is reversed when recovery of even a large part of the message is desirable. If the base can afford to dispatch a large cluster of drones, the increased receive diversity will make the system resilient to packet erasures but performance degradation will be abrupt beyond an erasure probability value, e.g. $\epsilon\approx 0.7$ for $L_1=8$.

\begin{figure}[h]
\centering
\includegraphics[width=8.65cm]{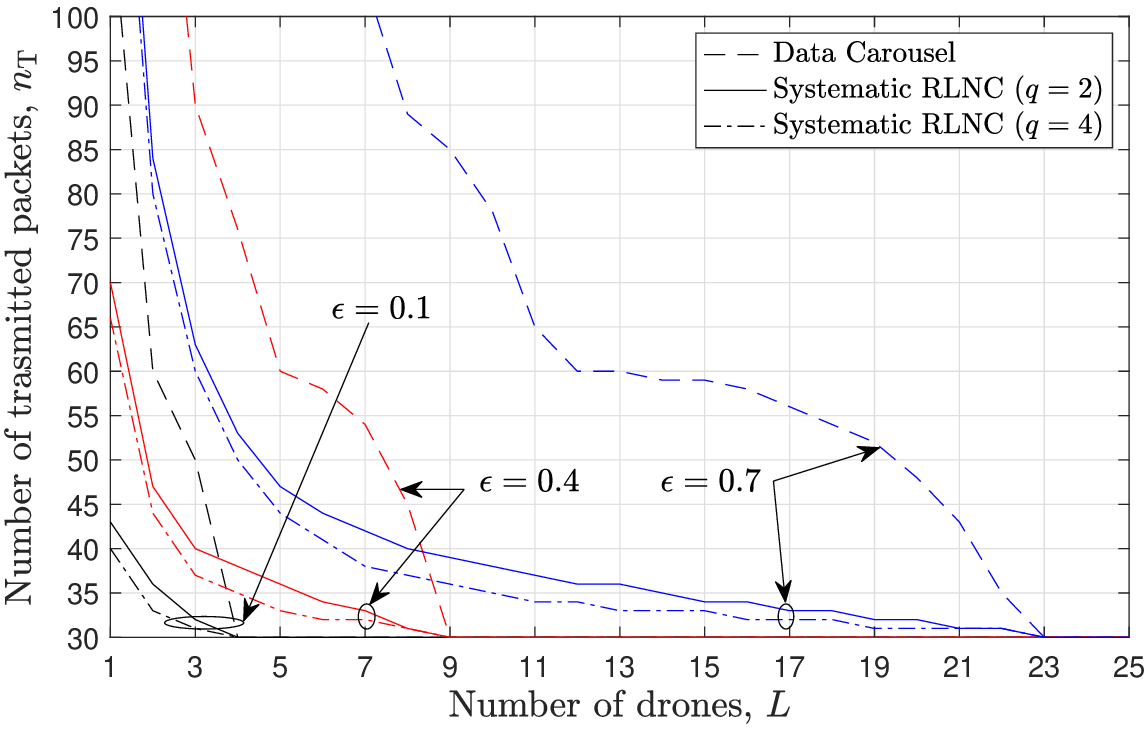}
\caption{Required number of packet transmissions ($n_\mathrm{T}$) as a function of the number of drones ($L$) for a single base (i.e., $N=1$) to fully decode a message of $k=30$ source packets with probability 99\%, when the packet erasure probability is $\epsilon\in\{0.1,0.4,0.7\}$.}
\label{fig:drones_vs_packets}
\vspace{4pt}
\end{figure}

When a single base ($N=1$) is required to collect data from a remote source node, which needs to keep packet transmissions as low as possible due to energy or time constraints, the optimal number of dispatched drones depends on the expected packet error probability and the chosen transmission method. As in the previous example, we assume that the links between the source node and the relay drones are characterized by the same average packet erasure probability, denoted by $\epsilon$. Fig.~\ref{fig:drones_vs_packets} depicts the impact of the number of drones $L$ on the number of transmitted packets $n_\mathrm{T}$ for $\epsilon\in\{0.1,0.4,0.7\}$, when the target probability of mission success is $99\%$. For a given value of $\epsilon$, we observe that there is a minimum value of $L$ for which the number of transmitted packets becomes equal to the number of source packets, which has been set to $k=30$. For instance, when $\epsilon=0.4$, no more than $L=9$ drones are required to reduce the equivalent erasure probability at the base, given by \eqref{eq.equiv_erasure_prob}, to a value that minimises the number of transmitted packets to $n_\mathrm{T}=30$. Furthermore, if $L=9$ drones are dispatched, the simple data carousel can be used. The additional encoding and decoding complexity introduced by systematic RLNC can be justified when the available drones are $L=8$ or fewer, as the value of $n_\mathrm{T}$ does not increase as sharply as in the case of the data carousel. If a low value of $n_\mathrm{T}$ is essential, an increase in the size of the field over which RLNC is performed, from $q=2$ to $q=4$, can marginally decrease the number of transmitted packets, as shown in Fig.~\ref{fig:drones_vs_packets}.

\vspace{8pt}
\section{Conclusions}
\label{sec:conclusions}

This paper considered a source node that broadcasts \mbox{packets} to clusters of drones using either a data carousel or systematic RLNC. Successfully received packets are stored, carried and delivered to bases, which can be either isolated or interconnected. A theoretical framework was developed for the calculation of (i) the probability that a particular base will retrieve some or all of the source packets, and (ii) the probability that all bases will obtain all of the source packets. The framework was validated through simulations and performance trade-offs were identified. Findings established that systematic RLNC over large finite fields should be used when full message recovery is desirable, whereas RLNC over small finite fields could be employed when partial message recovery is essential. Even though systematic RLNC offers a clear advantage over a data carousel, in terms of data reliability, the latter transmission method is still a viable solution if simplicity in the communication process takes priority over the cost of having a large number of drones in each cluster.
\balance

\bibliographystyle{IEEEtran}
\bibliography{IEEEabrv,IEEE_references}

\end{document}